\documentclass{article}

\usepackage{graphicx}

\usepackage{natbib}
\usepackage{journals_joge}   

\begin{document}

\title{Free core nutation: new large disturbance and connection evidence with geomagnetic
jerks\footnote{Acta Geodyn. Geomater., 2016, v. 13, No. 1, 41-45. DOI: 10.13168/AGG.2015.0042}}
\author{Zinovy Malkin$^{1,2,3}$\\
$^1$Pulkovo Observatory, St. Petersburg 196140, Russia \\
$^2$St. Petersburg State University, St. Petersburg 199034, Russia\\
$^3$Kazan Federal University, Kazan 420000, Russia\\
e-mail: malkin@gao.spb.ru}
\date{~}
\maketitle

\begin{abstract}
Variations in free core nutation (FCN) are connected with various processes in the Earth's fluid core and core-mantle coupling,
which are also largely responsible for the geomagnetic field variations, particularly the geomagnetic jerks (GMJs).
A previous study (Malkin, 2013) revealed that the epochs of the observed extremes in the FCN amplitude and phase variations are close to the GMJ epochs.
In this paper, a new evidence of this connection was found.
The large FCN amplitude and phase disturbance occurred at the epoch close to the newly revealed GMJ 2011.
This event occurred to be the second large change in the FCN amplitude and phase after the 1999 disturbance that is also associated
with the GMJ 1999. 
Moreover, the long-time FCN phase drift had changed suddenly in 1998--1999, immediately before the GMJ 1999, and seemed to change again
at the epoch immediately preceding the GMJ 2011.
The FCN amplitude showed a general long-time decrease before GMJ 1999, and it subsequently grew until GMJ 2011, and then seemed to decrease again.
A smaller FCN change can be observed at the epoch around 2013, which is also suspected as the GMJ epoch.
The latter confirms the suggestion that a rapid change in the FCN amplitude and/or phase can be used as an evidence of the GMJ that is not clearly
detected from the geomagnetic observations. 
\end{abstract}



\section{Introduction}

Retrograde free core nutation (FCN) is a component of the nutation motion of the Earth's rotational axis in space.
FCN causes variations in the position of the celestial pole, with period of about 430 solar days and amplitude of approximately 0.15--0.2~mas.
Investigation of the FCN is important to improve the theoretical modeling of the Earth rotation and better understand the Earth's interior.
The FCN amplitude and phase significantly vary with time, and its excitation mechanism has not yet been fully elucidated.
It is supposed that the FCN is mainly excited by the atmosphere with ocean contribution \citep{Dehant2003a,Brzezinski2005,Lambert2006,Vondrak2014}. 
However, this mechanism cannot explain all the details of the FCN amplitude and especially phase variations.

Another source of FCN excitation can be the geomagnetic field (GMF) variations, particularly the geomagnetic jerks (GMJs),
which are observed as rapid changes in the GMF secular variations.
They occur on a time scale of about one year, 1--2 times per decade, and are registered at geomagnetic observatories, as well as during
the recent years from satellite observations by using the measurements of the vertical and horizontal components of the GMF, magnetic declination, etc.

Our previous study \citep{Malkin2013d} showed that the observed extremes in the FCN amplitude and phase variations, as well as
their derivatives, are close to the GMJ epochs.
It can tell us that the FCN can be excited by the same processes that cause the GMJs. 
This assumption seems to be close to reality because the GMFs are mostly generated by the flows in the core, and the same flows lead to variations
of the core moments of inertia (as well as, to a lesser extent, the whole Earth), and thus can cause the FCN variations \citep{Dehant2003b}.

\citet{Vondrak2014} independently confirmed the connection between GMJs and FCN.
The authors investigated the excitation of the nutation motion of the Earth's spin axis in space, including FCN, by the atmosphere and ocean.
Particularly, they compared the series of geophysical excitations with the observed nutation angles by using numerical integration of the
\citet{Brzezinski1994} broad-band Liouville second-order differential equations.
The authors found that applying re-initialization of the integration at epochs of GMJs substantially improves the agreement between
the integrated and observed nutation angles.

This paper is a continuation of the previous work \citep{Malkin2013d}, which was extended in two respects.
First, a longer time span of observations was used, which allowed us to confront the astrometric observations made during the recent years with newly
revealed GMJs.
Second, an in-depth analysis was performed including the first and second derivatives of both FCN amplitude and phase variations.
In result, a new sudden change in the FCN amplitude and phase was revealed.
This sudden change is the second large change after a similar event in 1999 that is associated with the GMJ 1999.
A smaller change in the FCN amplitude and phase was detected at the epoch around 2013, which is also suspected as a GMJ epoch.

The paper is organized as follows.
In Section~\ref{sect:results}, the FCN model is described which was used in this study, and the FCN amplitude and phase variations are computed
and confronted with the GMJs.
In the concluding Section~\ref{sect:conclusion}, the results obtained in the previous section are discussed.


\section{FCN model and GMJs}
\label{sect:results}

All FCN models are constructed based on the analysis of the celestial pole offset (CPO) time series obtained from
very long baseline interferometry (VLBI) observations of extragalactic radio sources.
CPOs are the differences $dX$ and $dY$ between the observed celestial pole position $X$ and $Y$ and the International Astronomical Union (IAU)
official celestial intermediate pole, which is currently modeled by the IAU 2000/2006 precession-nutation theory \citep{IERSConv2010}.

\begin{figure*}[t]
\centering
\includegraphics[width=\textwidth,clip]{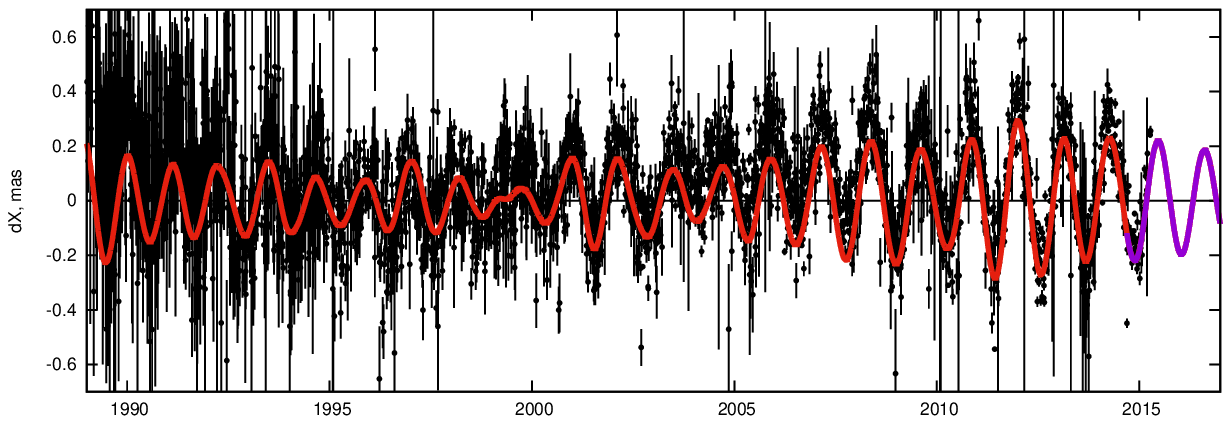} \\
\includegraphics[width=\textwidth,clip]{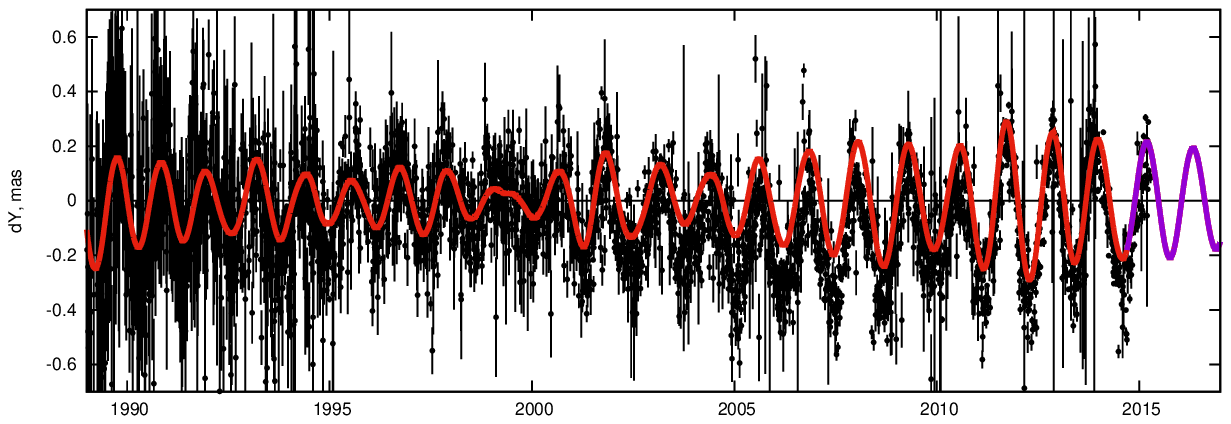}
\caption{IVS CPO series. Each point corresponds to one VLBI observing session.
Red line represents the ZM3 FCN model, and FCN prediction is shown in purple.}
\label{fig:cpo_ivs}
\end{figure*}

Figure~\ref{fig:cpo_ivs} shows the combined CPO series provided by the International VLBI Service for Geodesy and Astrometry (IVS),
\citep{Boeckmann2010,Schuh2012}.
The CPO data comprise two principal components, namely, the (quasi) periodic FCN term with a period of approximately 430 solar days and an average
amplitude of about 0.2~mas, as well as low-frequency changes, including trend and long-period harmonics of similar amplitude caused mainly by
the inaccuracy of the precession-nutation model.
One can see in the figure that the VLBI data obtained before the 1990s are very noisy and have relatively large uncertainties.
However, in this paper, the FCN amplitude and phase variations during the years after 2007 are mainly considered.
Earlier data were analyzed in detail in \citet{Malkin2013d}. 

Three FCN models are currently available for scientific and practical users:
two models of the author, namely ZM1 \citep{Malkin2004i} and ZM3 \citep{Malkin2013d} that are available at the Pulkovo Observatory
website\footnote{http://www.gao.spb.ru/english/as/persac/},
and the Lambert model \citep{IERSConv2010} that is available at the Paris Observatory website\footnote{http://syrte.obspm.fr/$\sim$lambert/fcn/}.
They are regularly updated, thus they always contain the most recent data.
These three models are compared in \citet{Malkin2007ie,Malkin2011h,Malkin2013d}.
Large differences were not observed between them for the investigation of FCN and GMJ interconnection.
The FCN amplitude and phase variation of the three FCN models are similar.
All three FCN series show the same epochs of the extreme FCN amplitude and phase, as well as their derivatives.

The most recent FCN model ZM3 \citep{Malkin2013d} was used for this study.
The parameters of this model were computed by running 431-day intervals (the nearest odd number of days to the FCN period) with one-day shift. 
At each interval, the four parameters were adjusted according to the following equations:
\begin{equation}
\begin{array}{rcl}
dX &=& A_c \cos \varphi - A_s \sin \varphi + X_0 \,, \\
dY &=& A_c \sin \varphi + A_s \cos \varphi + Y_0 \,,
\end{array}
\label{eq:SL}
\end{equation}
where $\varphi = 2\pi / P_{FCN} (t-t_0)$, $P_{FCN}$ is the FCN period equal to --430.21 solar days recommended by the IERS
Conventions (2010) \citep{IERSConv2010}, $t_0$=J2000.0, and $t$ is the epoch at which the $dX$ and $dY$ values are given.
Each pair in Eq.~(\ref{eq:SL}) corresponds to one CPO epoch given with one-day step.
The model parameters $A_c$, $A_s$, $X_0$, and $Y_0$ were computed at the middle epoch of each 431-day interval.
Thus, the resulting FCN parameter table is given with one-day step.
The FCN contribution to the celestial pole motion at the given epoch was computed by using Eq.~(\ref{eq:SL}) without the shift terms $X_0$ and $Y_0$.
The $A_c$ and $A_s$ parameters should be interpolated at the required date.
For daily parameter table, the linear interpolation provides sufficient accuracy. 

\begin{figure*}
\centering
\includegraphics[width=0.48\textwidth,clip]{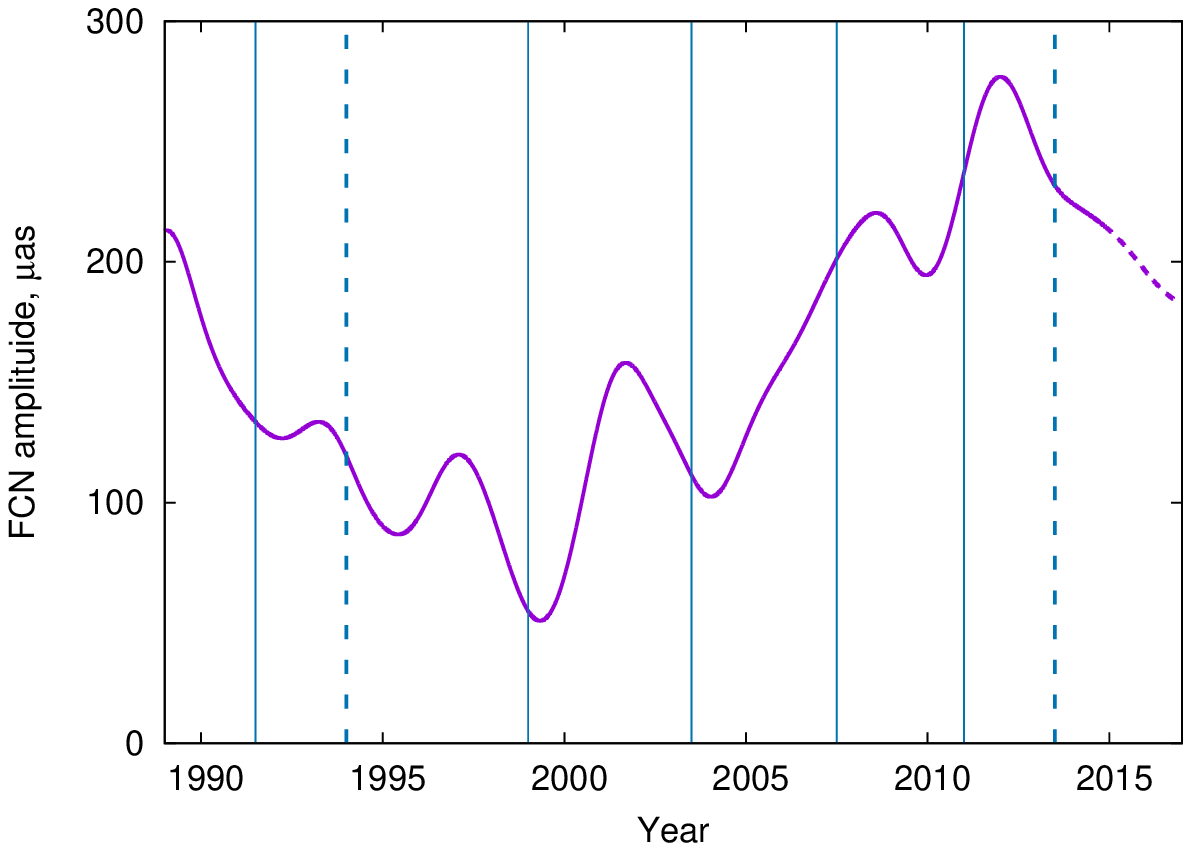}  \hfill \includegraphics[width=0.48\textwidth,clip]{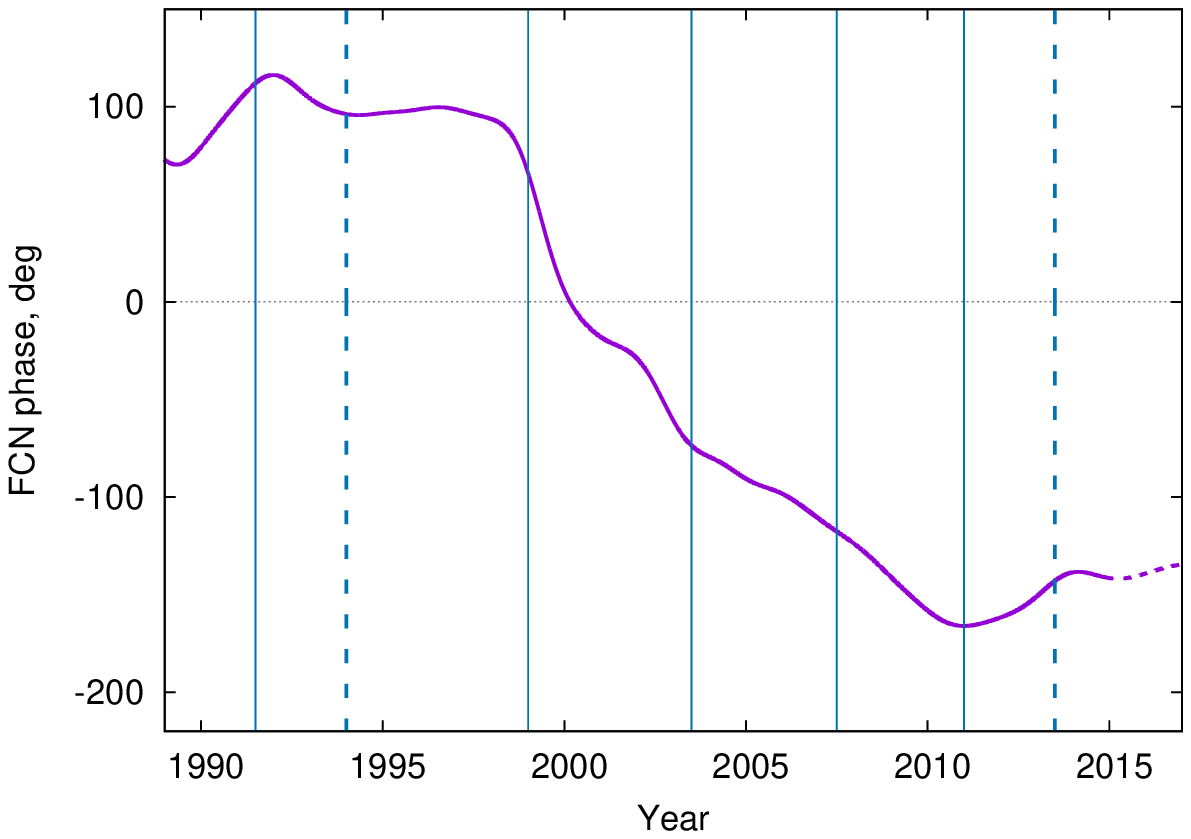} \\
\includegraphics[width=0.48\textwidth,clip]{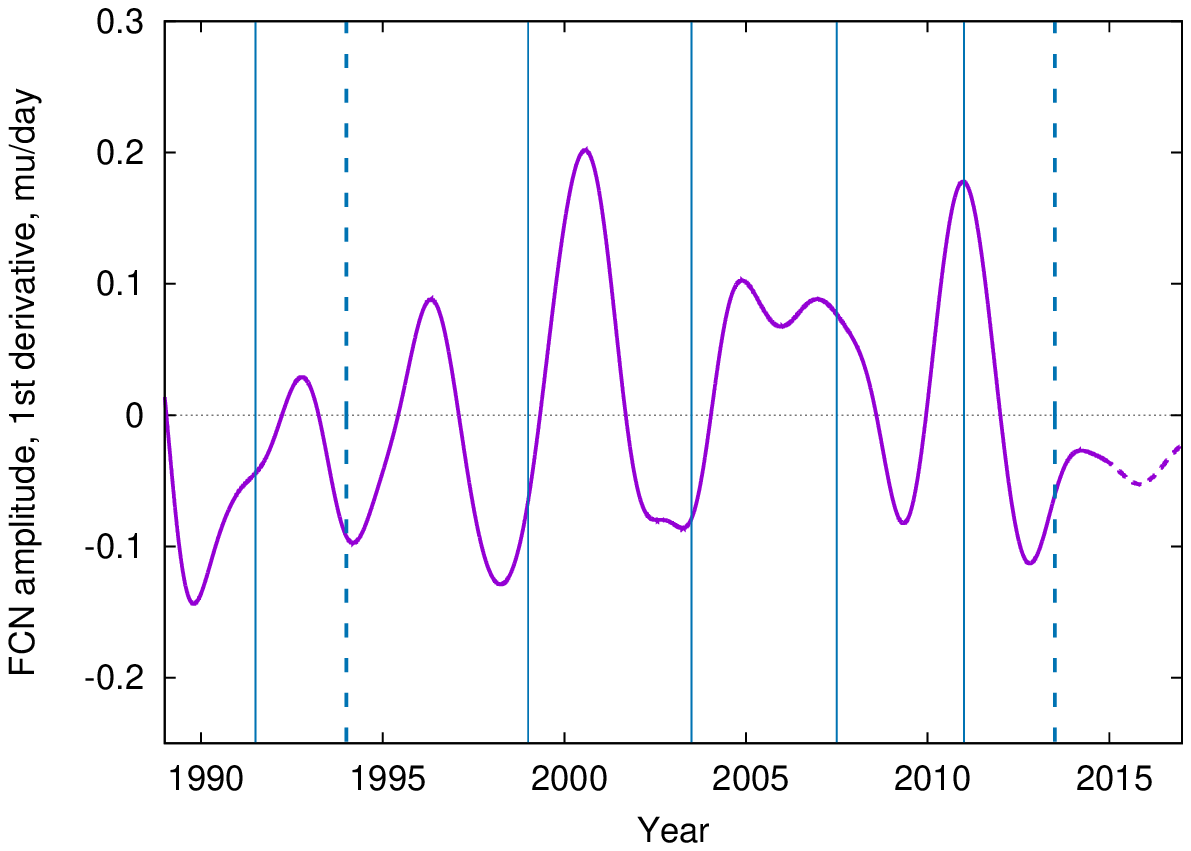} \hfill \includegraphics[width=0.48\textwidth,clip]{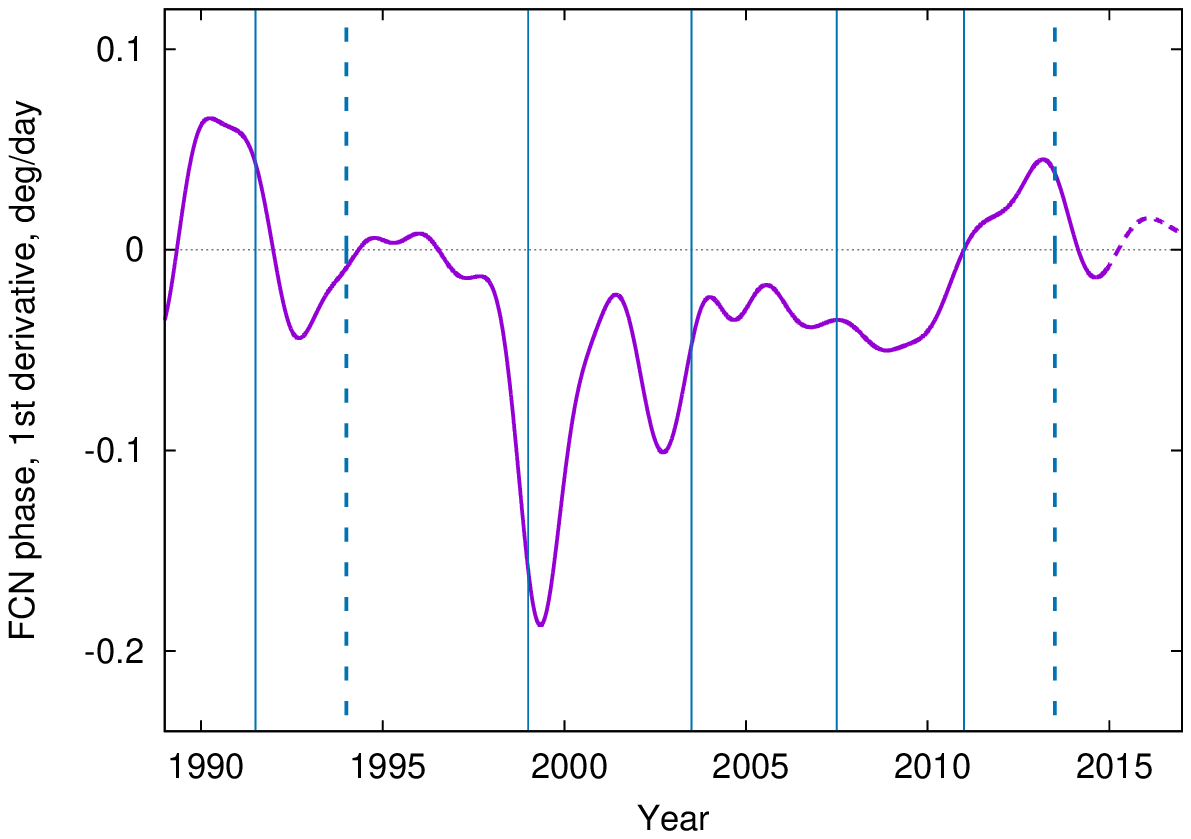} \\
\includegraphics[width=0.48\textwidth,clip]{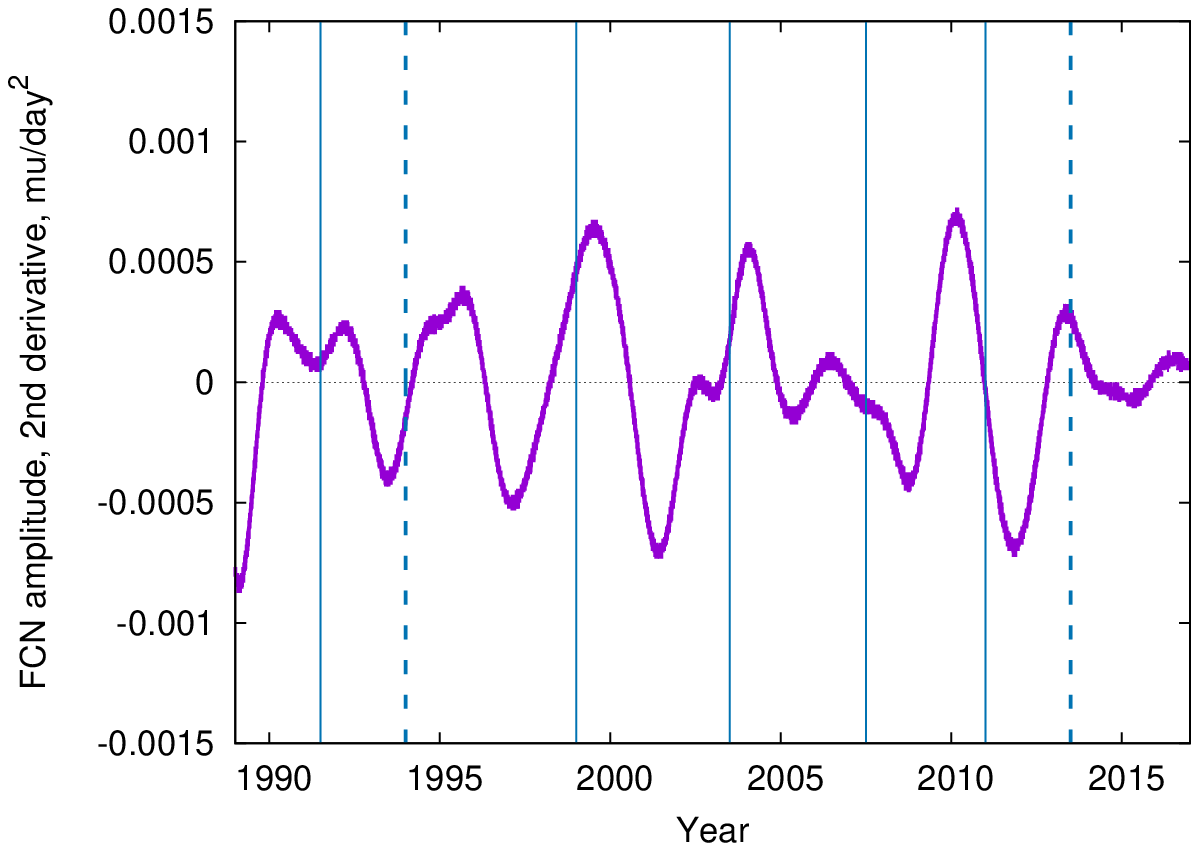} \hfill \includegraphics[width=0.48\textwidth,clip]{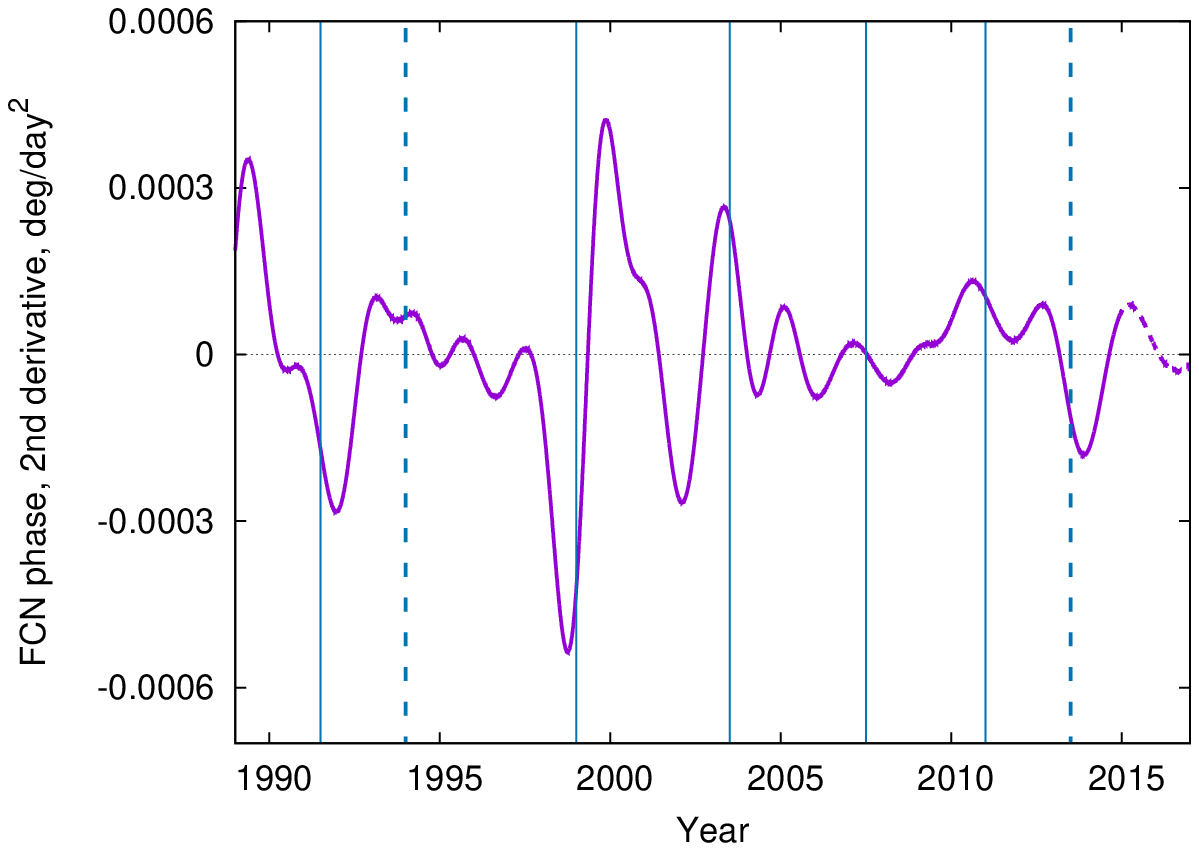}
\caption{Variations of the FCN amplitude (left) and phase (with the linear trend removed, right), as well as their first and second derivatives.
Prediction is shown by a dashed line. GMJs are shown with vertical lines, jerks that are not finally confirmed are shown with a dashed vertical line.}
\label{fig:fcn_jerks}
\end{figure*}

Figure~\ref{fig:cpo_ivs} depicts the FCN series computed with the model described above.
The FCN amplitude and phase variations with their first and second derivatives are presented in Fig.~\ref{fig:fcn_jerks}.
The linear trend corresponding to the nominal FCN period of $P_{FCN}$ is removed from the FCN phase series.

Figure~\ref{fig:fcn_jerks} shows also the GMJs observed after 1990.
These jerks were registered in 1991, 1999, 2003--2004, and 2007--2008 \citep{Mandea2010,Silva2012a,Chulliat2010,Kotze2011}.
Following \citet{Malkin2013d}, a possible jerk in 1994 was also added, which can be observed in the geomagnetic data obtained at several observatories
\citep{Nagao2002,Olsen2007,Mandea2010,PavonCarrasco2013}.
\citet{Chulliat2014} detected the 2011 jerk and quasi-definitive data for 2013, which may be a separate jerk or manifestation of the jerk 2011.
The supposed GMJs 1994 and 2013 that are not reliably detected from the geomagnetic observations are shown with a dashed line
in Fig.~\ref{fig:fcn_jerks}.


\section{Discussion}
\label{sect:conclusion}

\citet{Malkin2013d} demonstrated that the changes in the FCN amplitude and phase are closely related to the GMF sudden disturbances.
Observed extremes in the FCN amplitude and phase variations occurred at the epochs close to the GMJs epochs.
All the minima of the FCN amplitude followed the corresponding GMJs with a delay of 1--3 years.
This result is consistent with the conclusion of \citet{Gibert1998} and \citet{Bellanger2002}, who found that the rapid changes in the Chandler
wobble phase follow the GMJ with a delay of 1--3 years.
All the maxima of the second derivative of the FCN amplitude also occurred near the GMJ epochs.
The correlation between the FCN variations and GMJs was verified for all five GMJs that occurred from 1991 to 2003,
including the supposed jerk in 1994, which is not yet widely accepted in literature.
The result of \citet{Malkin2013d} was confirmed in this study, see Fig.~\ref{fig:fcn_jerks}, which corresponds to Fig.~5 of \citet{Malkin2013d}
for the interval before 2010.

In this paper, a new evidence was found of the connection between the FCN amplitude and phase variations and GMJs.
The substantial FCN amplitude and phase disturbance occurred at the epoch close to the newly revealed GMJ 2011.
Remarkably, this change in the FCN phase occurred to be the second large change after a similar event in 1999 that is
also associated with the GMJ 1999 and was detected for the first time by \citet{Shirai2005}.
The long-time FCN phase drift had changed suddenly in 1998--1999 immediately before the GMJ 1999, and seemed to change again
at the epoch immediately preceding the GMJ 2011.
A similar FCN amplitude behavior can be observed.
The FCN amplitude showed a general long-time decrease between 1990 and about GMJ 1999, and it subsequently grew with small variations until
GMJ 2011 and then seemed to decrease again. 

A smaller FCN change can be observed at the epoch around 2013, which is also suspected as the GMJ epoch. 
The latter confirms the suggestion made by \citet{Malkin2013d} that a rapid change in the FCN amplitude and/or phase can be used as
a supplement evidence of the GMJ that is not clearly detected from the geomagnetic observations.

\section*{Addendum}

After this paper was accepted for publication, the paper of \citet{Torta2015} appeared, in which the
GMJ in yearly 2014 was detected from a analysis of the GMF variations.
Thus, a GMJ around 2013 supposed from the FCN variations analysis performed in this paper is reaffirmed.

\section*{Acknowledgements}

The work was supported by the Russian Government Program of Competitive Growth of Kazan Federal University.
Critical comments and valuable suggestions of the anonymous referees are highly appreciated.

\bibliographystyle{joge}
\bibliography{my_eng,geodesy,astronomy,earth}

\begin{thebibliography}{25}
\providecommand{\natexlab}[1]{#1}
\providecommand{\doi}[1]{doi:\discretionary{}{}{}#1}
\providecommand{\url}[1]{{#1}}
\providecommand{\eprint}[2][]{\url{#2}}

\bibitem[{{Bellanger} et~al.(2002){Bellanger}, {Gibert}, and {Le
  Mou{\"e}l}}]{Bellanger2002}
{Bellanger} E, {Gibert} D, {Le Mou{\"e}l} JL (2002) {A geomagnetic triggering
  of Chandler wobble phase jumps?} \grl 29:1124. \doi{10.1029/2001GL014253}

\bibitem[{{B{\"o}ckmann} et~al.(2010){B{\"o}ckmann}, {Artz}, {Nothnagel}, and
  {Tesmer}}]{Boeckmann2010}
{B{\"o}ckmann} S, {Artz} T, {Nothnagel} A, {Tesmer} V (2010) {International
  {VLBI} {Service} for {Geodesy} and {Astrometry}: {Earth} orientation
  parameter combination methodology and quality of the combined products}.
  Journal of Geophysical Research (Solid Earth) 115:B04404.
  \doi{10.1029/2009JB006465}

\bibitem[{{Brzezi{\'n}ski}(1994)}]{Brzezinski1994}
{Brzezi{\'n}ski} A (1994) {Polar motion excitation by variations of the
  effective angular momentum function, II: extended model}. manuscripta
  geodaetica 19:157--171

\bibitem[{{Brzezi{\'n}ski}(2005)}]{Brzezinski2005}
{Brzezi{\'n}ski} A (2005) {Chandler wobble and free core nutation: observation,
  modeling and geophysical interpretation}. Artificial Satellites 40:21--33

\bibitem[{{Chulliat} and {Maus}(2014)}]{Chulliat2014}
{Chulliat} A, {Maus} S (2014) {Geomagnetic secular acceleration, jerks, and a
  localized standing wave at the core surface from 2000 to 2010}. Journal of
  Geophysical Research (Solid Earth) 119:1531--1543. \doi{10.1002/2013JB010604}

\bibitem[{{Chulliat} et~al.(2010){Chulliat}, {Th{\'e}bault}, and
  {Hulot}}]{Chulliat2010}
{Chulliat} A, {Th{\'e}bault} E, {Hulot} G (2010) {Core field acceleration pulse
  as a common cause of the 2003 and 2007 geomagnetic jerks}. \grl 37:L07301.
  \doi{10.1029/2009GL042019}

\bibitem[{{Dehant} and {Mathews}(2003)}]{Dehant2003b}
{Dehant} V, {Mathews} PM (2003) {Information about the core from Earth
  nutation}. In: {Dehant} V, {Creager} VKC, {Karato} S, {Zatman} S (eds)
  Earth's Core: Dynamics, Structure, Rotation, Geodyn. Ser., vol. 31, AGU,
  Washington, D. C., pp 263--277, \doi{doi:10.1029/GD031p0263}

\bibitem[{{Dehant} et~al.(2003){Dehant}, {Feissel-Vernier}, {de Viron}, {Ma},
  {Yseboodt}, and {Bizouard}}]{Dehant2003a}
{Dehant} V, {Feissel-Vernier} M, {de Viron} O, {Ma} C, {Yseboodt} M, {Bizouard}
  C (2003) {Remaining error sources in the nutation at the submilliarc second
  level}. Journal of Geophysical Research (Solid Earth) 108:2275.
  \doi{10.1029/2002JB001763}

\bibitem[{{Gibert} et~al.(1998){Gibert}, {Holschneider}, and {Le
  Mou{\"e}l}}]{Gibert1998}
{Gibert} D, {Holschneider} M, {Le Mou{\"e}l} JL (1998) {Wavelet analysis of the
  Chandler wobble}. \jgr 103:27,069--27,089. \doi{10.1029/98JB02527}

\bibitem[{{Kotz\'e} et~al.(2011){Kotz\'e}, {Korte}, and {Mandea}}]{Kotze2011}
{Kotz\'e} P, {Korte} M, {Mandea} M (2011) {Polynomial Modelling of Southern
  African Secular Variation Observations Since 2005}. Data Science Journal
  10:95--101

\bibitem[{{Lambert}(2006)}]{Lambert2006}
{Lambert} SB (2006) {Atmospheric excitation of the Earth's free core nutation}.
  \aap 457:717--720. \doi{10.1051/0004-6361:20065813}

\bibitem[{{Malkin}(2007)}]{Malkin2007ie}
{Malkin} ZM (2007) Empiric models of the {Earth's} free core nutation. \ssrr
  41:492--497. \doi{10.1134/S0038094607060044}

\bibitem[{{Malkin}(2004)}]{Malkin2004i}
{Malkin} Z (2004) Comparison of {VLBI} nutation series with the {IAU2000A}
  model. In: {Finkelstein} A, {Capitaine} N (eds) Proc. Journ\'ees 2003
  Syst\`emes de R\'ef\'erence Spatio-temporels, St. Petersburg, Russia, Sep
  22-25, pp 24--31

\bibitem[{{Malkin}(2011)}]{Malkin2011h}
{Malkin} Z (2011) Comparison of {CPO} and {FCN} empirical models. In:
  {Capitaine} N (ed) Proc. Journ\'ees 2010 Syst\`emes de R\'ef\'erence
  Spatio-temporels, Observatoire de Paris, 20--22 Sep, pp 172--175

\bibitem[{{Malkin}(2013)}]{Malkin2013d}
{Malkin} Z (2013) {Free core nutation and geomagnetic jerks}. Journal of
  Geodynamics 72:53--58. \doi{10.1016/j.jog.2013.06.001}

\bibitem[{{Mandea} et~al.(2010){Mandea}, {Holme}, {Pais}, {Pinheiro},
  {Jackson}, and {Verbanac}}]{Mandea2010}
{Mandea} M, {Holme} R, {Pais} A, {Pinheiro} K, {Jackson} A, {Verbanac} G (2010)
  {Geomagnetic Jerks: Rapid Core Field Variations and Core Dynamics}. \ssr
  155:147--175. \doi{10.1007/s11214-010-9663-x}

\bibitem[{{Nagao} et~al.(2002){Nagao}, {Higuchi}, {Iyemori}, and
  {Araki}}]{Nagao2002}
{Nagao} H, {Higuchi} T, {Iyemori} T, {Araki} T (2002) {Automatic detection of
  geomagnetic jerks by applying a statistical time series model to geomagnetic
  monthly means}. Progresses in Discovery Science, Lecture Notes in Computer
  Science 2281:360--371

\bibitem[{{Olsen} and {Mandea}(2007)}]{Olsen2007}
{Olsen} N, {Mandea} M (2007) {Investigation of a secular variation impulse
  using satellite data: The 2003 geomagnetic jerk}. \epsl 255:94--105.
  \doi{10.1016/j.epsl.2006.12.008}

\bibitem[{{Pav{\'o}n-Carrasco} et~al.(2013){Pav{\'o}n-Carrasco}, {Torta},
  {Catal{\'a}n}, {Talarn}, and {Ishihara}}]{PavonCarrasco2013}
{Pav{\'o}n-Carrasco} FJ, {Torta} JM, {Catal{\'a}n} M, {Talarn} {\`A},
  {Ishihara} T (2013) {Improving total field geomagnetic secular variation
  modeling from a new set of cross-over marine data}. \pepi 216:21--31.
  \doi{10.1016/j.pepi.2013.01.002}

\bibitem[{{Petit} and {Luzum}(2010)}]{IERSConv2010}
{Petit} G, {Luzum} B (eds)  (2010) IERS Conventions (2010). IERS Technical Note
  No.~36, Verlag des Bundesamts f\"ur Kartographie und Geod\"asie, Frankfurt am
  Main

\bibitem[{{Schuh} and {Behrend}(2012)}]{Schuh2012}
{Schuh} H, {Behrend} D (2012) {VLBI: A fascinating technique for geodesy and
  astrometry}. \jog 61:68--80. \doi{10.1016/j.jog.2012.07.007}

\bibitem[{{Shirai} et~al.(2005){Shirai}, {Fukushima}, and
  {Malkin}}]{Shirai2005}
{Shirai} T, {Fukushima} T, {Malkin} Z (2005) Detection of phase disturbances of
  free core nutation of the {Earth} and their concurrence with geomagnetic
  jerks. \eps 57:151--155

\bibitem[{{Silva} and {Hulot}(2012)}]{Silva2012a}
{Silva} L, {Hulot} G (2012) {Investigating the 2003 geomagnetic jerk by
  simultaneous inversion of the secular variation and acceleration for both the
  core flow and its acceleration}. \pepi 198:28--50.
  \doi{10.1016/j.pepi.2012.03.002}

\bibitem[{{Torta} et~al.(2015){Torta}, {Pav{\'o}n-Carrasco}, {Marsal}, and
  {Finlay}}]{Torta2015}
{Torta} JM, {Pav{\'o}n-Carrasco} FJ, {Marsal} S, {Finlay} CC (2015) {Evidence
  for a new geomagnetic jerk in 2014}. \grl 42:7933--7940.
  \doi{10.1002/2015GL065501}

\bibitem[{{Vondr{\'a}k} and {Ron}(2014)}]{Vondrak2014}
{Vondr{\'a}k} J, {Ron} C (2014) {Geophysical excitation of nutation ---
  comparison of different models}. Acta Geodyn Geomater 11:193--200

\end{thebibliography}

\end{document}